%% file: main.tex
\documentclass[sigconf,nonacm]{acmart}

\usepackage{graphicx}      
\usepackage{subcaption}    
\usepackage{amsmath}       
\usepackage{algorithm}     
\usepackage{algpseudocode} 
\usepackage{booktabs}      
\usepackage{multirow}      

\renewcommand\footnotetextcopyrightpermission[1]{}
\title{Ising Acceleration for Multi-Robot Multi-Target Planning}

\author{Ahmet Efe, Recep B. Uludag, Chris H. Kim, and Ulya R. Karpuzcu}

\affiliation{%
  \institution{University of Minnesota}
  \city{Minneapolis}
  \state{Minnesota}
  \country{USA}
}

\begin{document}

\begin{abstract}
Ising machines are emerging as promising hardware for combinatorial
optimization. With recent advances in CMOS Ising technology, they are
becoming attractive as low-power accelerator systems for robotics, where
energy is limited and combinatorial optimization arises in multiple
forms. However, a hardware-aware analysis of where such chips fit within
a robotics planning stack is still missing. This paper studies the
capabilities and limitations of CMOS Ising machines for low-power
acceleration in multi-robot multi-target planning.

We analyze three planning layers---target sharing, tour construction,
and pathfinding---using real 45-spin all-to-all connected CMOS Ising
chips as representative devices. We propose new Ising-based planning
methods and a multi-mapping pipeline that uses spin merging, coefficient
quantization, and spin-budget branching to adapt subproblems to
spin- and coefficient-limited hardware. Our results show that the
proposed recursive target-sharing method naturally matches the Ising
hardware, achieving up to 8{,}000$\times$ lower energy than a classical
baseline. End to end, the Ising
pipeline produces routes within 9\% of a strong classical baseline at
130$\times$ lower energy, showing that compact CMOS Ising machines can
be effective in selected parts of the planning stack.
\end{abstract}

\maketitle
\linespread{0.90}

\section{Introduction}
\label{sec:introduction}
\input{sec/1_introduction}

\section{Background}
\label{sec:background}
\input{sec/2_background}

\section{Workload Characterization}
\label{sec:workload}
\input{sec/3_workload}

\section{Hardware-Aware Ising Mapping Pipeline}
\label{sec:mapping_pipeline}
\input{sec/4_mapping_pipeline}

\section{Case Studies}
\label{sec:case_studies}
\input{sec/5_case_studies}

\section{Evaluation}
\label{sec:evaluation}
\input{sec/6_evaluation}

\section{Related Work}
\label{sec:related_work}
\input{sec/7_related_work}

\section{Conclusion}
\label{sec:conclusion}
\input{sec/8_conclusion}

\bibliographystyle{ACM-Reference-Format}
\bibliography{ref}

\makeatletter
\let\baselinestretch\ACM@origbaselinestretch
\makeatother

\end{document}

%% file: sec/1_introduction.tex
Robotic systems that operate autonomously---warehouse robots, delivery
drones, and search-and-rescue teams---must make a stream of combinatorial
planning decisions under tight energy budgets. A multi-robot fleet
visiting a set of targets in an obstacle-filled environment faces three
interleaved problems: assigning targets to robots, ordering each robot's
assigned targets, and finding obstacle-avoiding paths between successive
waypoints.

Classical solvers for these problems are mature and effective, but they
run on general-purpose processors whose energy cost can be significant
on battery-powered platforms. This has motivated interest in specialized
hardware such as Ising machines, which approximately minimize quadratic
binary objectives by exploiting device dynamics. Recent advances in
CMOS-based Ising chips have produced small, low-power devices that can
be integrated alongside a host processor, making them attractive
candidates for on-board robotics acceleration.

However, it is not clear whether current Ising hardware can handle
realistic robotics workloads. Today's compact CMOS Ising chips offer
limited spin counts, narrow coefficient ranges, and fixed hardware
interfaces. The question is not whether robotics problems \emph{can} be
expressed as Ising optimization in principle---they can---but whether
the resulting models can be mapped to real chips at useful problem
sizes, and whether the energy savings justify the decomposition and
mapping effort. What is missing is a hardware-aware, end-to-end analysis
of where compact low-power CMOS Ising chips fit within a robotics
planning stack.

This paper studies that question using multi-robot multi-target planning
as both a target application and a structured workload. The problem
naturally decomposes into three planning layers---target sharing, tour
construction, and pathfinding---that stress different hardware limits.
For each layer, we develop a hardware-aware Ising-based method:
recursive target sharing as a distance-weighted binary partitioning
problem, clustered tour construction using spin-limited Ising
subproblems, and patch-sliding pathfinding using local Ising solves.
Together, these methods show where compact CMOS Ising hardware provides
useful acceleration, where coefficient and spin limits break direct
mappings, and which mapping strategies recover useful candidates.

\begin{figure}[t]
    \centering
    \includegraphics[
        width=1\columnwidth,
        trim={1cm 0.5cm 1cm 1cm},
        clip
    ]{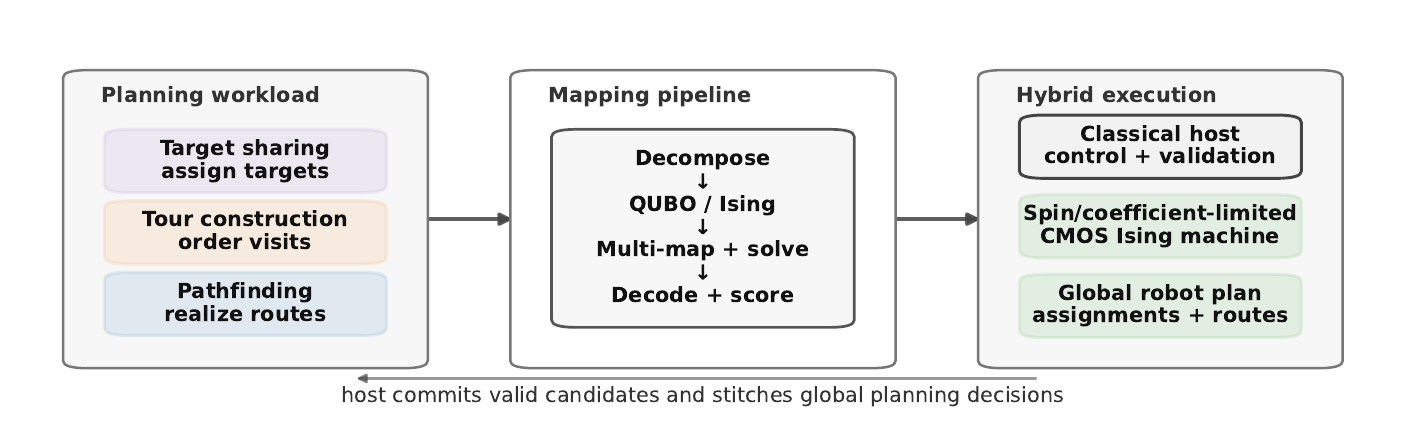}
    \vspace{-0.4cm}
    \caption{Hardware-aware Ising acceleration pipeline. A spin- and
    coefficient-limited CMOS Ising chip generates candidates, while the
    host validates, scores, and stitches them into a global robot plan.}
    \label{fig:paper_overview}
    \vspace{-0.4cm}
\end{figure}

Rather than attempting to solve any planning layer globally on the chip,
we decompose each layer into chip-sized subproblems and generate multiple
hardware-compatible mappings per subproblem. The chip acts as a
low-power candidate-generation accelerator, not as a standalone planner.
Classical logic does not repair Ising outputs; it validates, scores, and
stitches candidates generated by the Ising solver.

Our contributions are as follows:
\begin{itemize}
\item We provide a hardware-aware characterization of multi-robot
multi-target planning on compact CMOS Ising hardware, showing that
target sharing, tour construction, and pathfinding expose different
bottlenecks: spin count, coefficient range, or decode validity.

\item We propose a multi-mapping pipeline for spin- and
coefficient-limited Ising accelerators. The pipeline decomposes planning
problems into chip-sized subproblems and evaluates mapping portfolios
formed by spin merging, coefficient quantization, and spin-budget
branching.

\item We propose three Ising-based planning methods for the MRMT stack:
patch-sliding pathfinding with spin merging, clustered tour construction
with spin-budget branching, and recursive target sharing with
coefficient quantization and batching. Classical logic is used for
decomposition, validation, scoring, and stitching rather than repair.

\item We evaluate the resulting Ising pipeline on a real 45-spin
all-to-all connected CMOS Ising chip. Pathfinding and target sharing run
directly on the chip, achieving energy reductions of 37$\times$ and
8{,}000$\times$, while tour construction exposes a coefficient-range
limitation. End to end, the Ising pipeline produces routes within 9\% of
a strong classical baseline at 130$\times$ lower energy.
\end{itemize}

The rest of the paper is organized as follows.
Section~\ref{sec:background} provides background on Ising machines and
multi-robot planning. Section~\ref{sec:workload} characterizes the three
planning layers and their QUBO formulations.
Section~\ref{sec:mapping_pipeline} describes the shared mapping
pipeline. Section~\ref{sec:case_studies} instantiates the pipeline for
each layer. Section~\ref{sec:evaluation} presents the experimental
evaluation. Sections~\ref{sec:related_work} and~\ref{sec:conclusion}
discuss related work and conclude.

%% file: sec/2_background.tex
\subsection{Ising Machines and QUBO}
\label{sec:background:ising}

A quadratic unconstrained binary optimization (QUBO) problem seeks a binary vector $x \in \{0,1\}^n$ that minimizes a quadratic objective
\begin{equation}
E_{\mathrm{QUBO}}(x) =
\sum_i q_i\, x_i +
\sum_{i<j} Q_{ij}\, x_i x_j,
\label{eq:qubo}
\end{equation}
where $q_i$ and $Q_{ij}$ are real-valued coefficients encoding the problem structure. A constant offset is omitted because it does not affect the optimizer. Many combinatorial optimization problems---including graph partitioning, satisfiability, and variants of the traveling salesman problem---can be cast in this form~\cite{lucas2014ising}.

An equivalent representation uses Ising spins $s_i \in \{-1,+1\}$ with the energy
\begin{equation}
E_{\mathrm{Ising}}(s) =
\sum_i h_i\, s_i +
\sum_{i<j} J_{ij}\, s_i s_j,
\label{eq:ising}
\end{equation}
where $h_i$ are local fields and $J_{ij}$ are pairwise couplings. The two forms are related by the substitution $x_i = (1+s_i)/2$. Ising machines are physical systems designed to approximately find low-energy spin configurations of~\eqref{eq:ising} by exploiting device dynamics such as annealing, oscillator coupling, or probabilistic bit flipping~\cite{mohseni2022ising}. Their appeal is that part of the optimization is carried out by the hardware dynamics rather than by sequential software, potentially offering energy and latency advantages over general-purpose processors.

\subsection{Ising Hardware Landscape}
\label{sec:background:hardware}

Ising machines span a wide range of scales and technologies. At the large end, quantum annealers such as D-Wave systems offer thousands of qubits but require cryogenic cooling and occupy rack-scale infrastructure~\cite{johnson2011dwave}. Coherent Ising machines (CIMs) based on optical parametric oscillators have been demonstrated with up to 100,000 spins, but they rely on fiber-loop cavities and laboratory-scale optical setups~\cite{honjo2021cim}. Digital annealers and simulated-bifurcation systems can also reach large spin counts through digital or time-multiplexed simulation~\cite{tatsumura2021scaling}. While these systems achieve impressive scale, they are not the target class for this work: quantum annealers and optical CIMs require infrastructure unsuitable for mobile robots, while large digital annealing systems target scale rather than tightly integrated low-power coprocessing.

At the other end of the spectrum, CMOS-based Ising chips target compact and low-power operation. Prior designs include CMOS annealing processors, fully connected digital annealers, coupled-oscillator chips, and probabilistic p-bit solvers~\cite{yamaoka2016ising,takemoto2020ising,yamamoto2021statica,aadit2022massively,aadit2024alltoall,cilasun2025cobi}. These systems show that Ising acceleration can be realized in several hardware styles, but their practical mapping constraints vary widely.

This paper focuses on compact all-to-all connected CMOS Ising chips that
can serve as plug-in coprocessors alongside a host CPU. In this class of
hardware, all-to-all connectivity simplifies embedding, but practical
mapping is shaped by several constraints. \textbf{Limited spin count}
means that application-scale problems cannot be mapped globally.
\textbf{Narrow coefficient ranges} limit the dynamic range available for
encoding constraint penalties and objective terms, making structures such
as one-hot permutation constraints difficult to realize. \textbf{Parallel
solver resources}, either within a chip or across multiple chips on the
same board, provide a natural way to evaluate multiple mappings.
\textbf{Low-power operation} makes these systems attractive for embedded
and battery-powered platforms.

We target representative hardware from this class: 45-spin all-to-all
connected CMOS Ising cores with narrow integer coefficient ranges,
parallel solver resources, and a low-power operating envelope. These
constraints motivate the hardware-aware analysis and mapping pipeline
developed in the rest of the paper, but the pipeline is not specific to
any single chip design.

\subsection{Multi-Robot Multi-Target Planning}
\label{sec:background:planning}

Multi-robot multi-target (MRMT) planning considers a team of $R$ robots
operating in a shared environment with obstacles. A set of $N$ target
locations must be collectively visited, and the planner must decide the
assignment of targets to robots, the visit order for each robot, and the
obstacle-aware paths between successive waypoints. This setting arises
in applications such as warehouse logistics, search and rescue, and
autonomous inspection.

The MRMT planning problem can be decomposed into three layers, each
corresponding to a different combinatorial subproblem:

\noindent\textbf{Target sharing.}
Deciding which robot is responsible for each target. This is an
allocation problem related to the multi-robot task allocation (MRTA)
literature~\cite{gerkey2004formal,korsah2013comprehensive} and to the
multiple traveling salesman problem (mTSP)~\cite{bektas2006multiple}.

\noindent\textbf{Tour construction.}
Given a robot and its assigned targets, deciding the order in which the
targets should be visited. This is an open-route variant of the
traveling salesman problem.

\noindent\textbf{Pathfinding.}
Given two locations and a map with obstacles, finding a valid low-cost
path. Classical algorithms such as A*~\cite{hart1968formal} and
Dijkstra's algorithm~\cite{dijkstra1959note} solve this efficiently on
general-purpose hardware.

These layers are coupled: obstacle-aware path distances are used by the
tour-construction and target-sharing layers, and target assignments
determine the per-robot tours that must be constructed. Section~\ref{sec:workload}
examines each layer as an Ising mapping target, and later sections
develop hardware-aware Ising methods for each layer.

%% file: sec/3_workload.tex
This section characterizes the three planning layers of multi-robot
multi-target planning as Ising mapping targets and shows why direct
QUBO/Ising formulations are insufficient on current compact CMOS Ising
hardware. The analysis motivates the hardware-aware mapping pipeline and
layer-specific Ising methods introduced in later sections.

\subsection{Pathfinding}
\label{sec:workload:pathfinding}

Pathfinding determines how a robot moves between two locations on a grid
map while avoiding obstacles. We view the grid as a graph $G=(V,E)$,
where each free cell is a node and edges connect horizontally or
vertically adjacent free cells. Given a start node $s$ and a goal node
$t$, the task is to select a connected sequence of neighboring nodes
from $s$ to $t$.

Several QUBO formulations of this problem are possible, differing in how
the path is represented.
A \emph{cell-based} formulation assigns one binary variable
$x_i \in \{0,1\}$ per free cell, indicating whether that cell belongs to
the path. The objective
\begin{equation}
E_{\mathrm{cell}} = A\, E_{\mathrm{end}} + A\, E_{\mathrm{conn}} + B \sum_i x_i
\label{eq:cell_path}
\end{equation}
enforces start/goal endpoints ($E_{\mathrm{end}}$), encourages
connectivity ($E_{\mathrm{conn}}$), and penalizes unnecessarily long
paths. This formulation scales as $O(|V|)$ and is the most compact: an
obstacle-free $10\times10$ grid requires 100~variables. However,
connectivity is only encouraged indirectly, so disconnected or looping
solutions can still appear.
An \emph{edge-based} formulation assigns a variable
$y_{ij}\in\{0,1\}$ to each possible move, scaling as $O(|E|)$ for
undirected edges or $O(2|E|)$ for directed edges. This represents motion
more directly but requires 180 undirected variables or 360 directed
variables on a $10\times10$ grid.
A \emph{time-expanded} formulation introduces
$z_{i,\tau}\in\{0,1\}$ for each node~$i$ and time step~$\tau$, encoding
the robot's position at every step. This is the most expressive model
but also the largest: a $10\times10$ grid over $T{=}20$ steps already
requires $100 \times 21 = 2{,}100$~variables.

All three direct full-map formulations exceed the 45-spin budget of the
target hardware for maps of practical size. Even the most compact
cell-based model requires 100~variables on a modest $10{\times}10$ grid.
Moreover, even when a local pathfinding instance is small enough in
variable count, endpoint and connectivity penalties can still produce
Ising coefficients outside the supported hardware range after
QUBO-to-Ising conversion.

\subsection{Tour Construction}
\label{sec:workload:tour}

\begin{figure}[t]
    \centering
    \includegraphics[width=0.7\columnwidth]{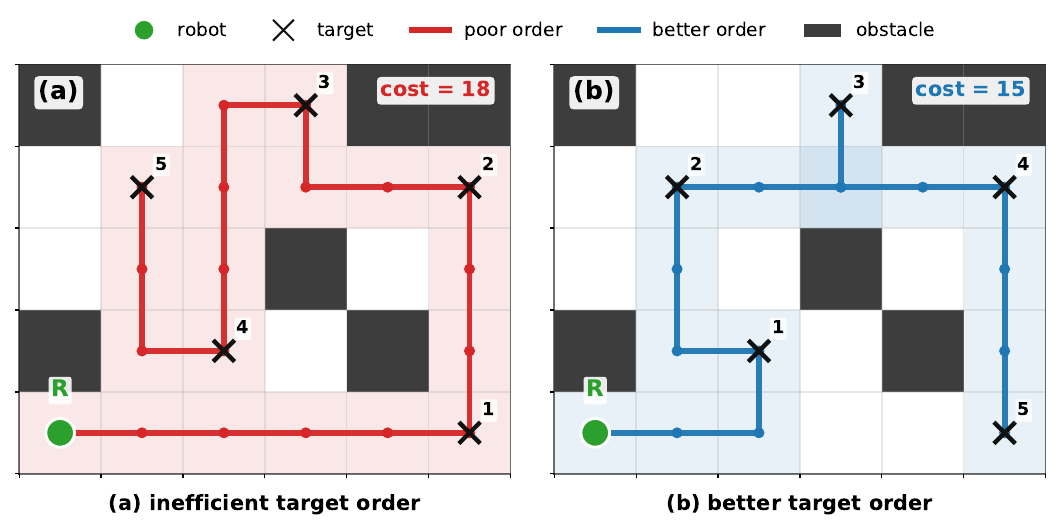}
    \vspace{-0.4cm}
    \caption{Tour construction as target ordering. The robot and target
    set are the same in both panels, but different visit orders produce
    different route costs. The tour solver therefore optimizes the order
    of targets using obstacle-aware path distances from the pathfinding
    layer.}
    \label{fig:tour_ordering_problem}
    \vspace{-0.4cm}
\end{figure}

When a robot must visit multiple targets, the planner must decide the
visit order. Given a start position and a target set
$\mathcal{T}=\{1,\ldots,N\}$, the goal is to find an open route---starting
at the robot and visiting every target exactly once---that minimizes
total path distance. The edge costs come from the pathfinding layer and
therefore reflect obstacles and map geometry.
Figure~\ref{fig:tour_ordering_problem} illustrates how different
orderings of the same target set produce different route costs.

This problem is closely related to the traveling salesman problem (TSP).
A standard QUBO formulation uses position-indexed variables
$x_{i,p}\in\{0,1\}$, where $x_{i,p}=1$ means target~$i$ is visited at
position~$p$. A valid tour corresponds to a permutation matrix, and the
objective
\begin{equation}
E_{\mathrm{TSP}} = A\, E_{\mathrm{target}} + A\, E_{\mathrm{position}} + B\, E_{\mathrm{route}}
\label{eq:tsp}
\end{equation}
penalizes repeated or missing targets ($E_{\mathrm{target}}$), empty or
overloaded positions ($E_{\mathrm{position}}$), and travel cost
($E_{\mathrm{route}}$).

The position-indexed formulation requires $N^2$ binary variables. A
7-target tour already needs $49$~variables, slightly above the 45-spin
hardware budget, while a 20-target tour needs $400$. However, the spin
count is not the only constraint. The one-hot row and column penalties
must dominate the route-cost terms to ensure valid permutations. On
hardware with a narrow integer coefficient range, these penalty
strengths are difficult to realize. If the penalties are too weak
relative to the route costs, the chip returns assignments with repeated
targets, missing positions, or other constraint violations. Thus, even a
small TSP instance that is close to the spin budget can be
\emph{numerically} incompatible with the hardware.

\subsection{Target Sharing}
\label{sec:workload:target_sharing}

\begin{figure}[t]
    \centering
    \includegraphics[width=0.7\columnwidth]{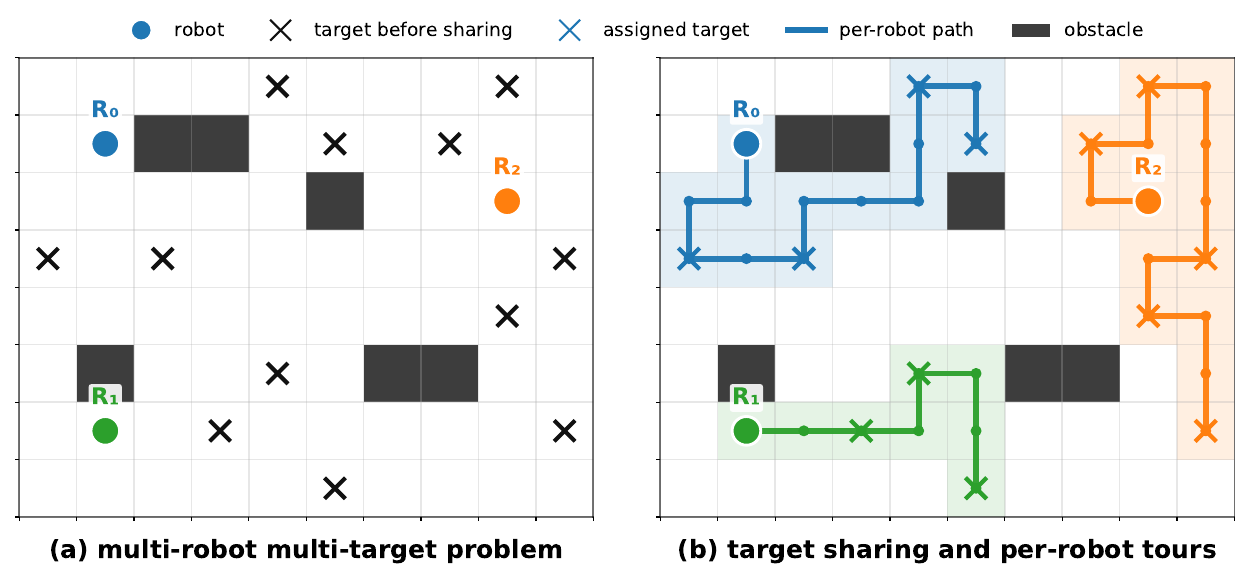}
    \vspace{-0.4cm}
    \caption{Multi-robot multi-target planning as target sharing plus
    per-robot tour construction. The left panel shows the full problem
    before allocation. The right panel shows a decomposed solution in
    which targets are assigned to robots and each robot constructs a
    route through its assigned targets.}
    \label{fig:target_sharing_decomposition}
    \vspace{-0.4cm}
\end{figure}

When multiple robots share a common set of targets, the planner must
decide which robot is responsible for each target. This is closely
related to the multiple traveling salesman problem~(mTSP), where the
planner makes two coupled decisions: target-to-robot assignment and
per-robot visit ordering. Even small instances create enormous search
spaces; three robots and twelve targets yield approximately
$12!\binom{14}{2} \approx 43$~billion possible assignment-and-ordering
combinations. Figure~\ref{fig:target_sharing_decomposition} illustrates
this decomposition.

A direct QUBO formulation assigns a variable
$x_{r,i,p}\in\{0,1\}$ for every robot~$r$, target~$i$, and tour
position~$p$:
\begin{equation}
E_{\mathrm{mTSP}} = A\, E_{\mathrm{visit}} + A\, E_{\mathrm{pos}} + B\, E_{\mathrm{route}},
\label{eq:mtsp}
\end{equation}
where $E_{\mathrm{visit}}$ ensures each target is visited exactly once,
$E_{\mathrm{pos}}$ enforces per-robot tour consistency, and
$E_{\mathrm{route}}$ captures travel cost. The variable count scales as
$RN^2$: three robots and twelve targets require
$3 \times 12^2 = 432$~variables, far beyond the hardware capacity. This
motivates our recursive target-sharing approach, which separates
allocation from ordering and replaces the global mTSP formulation with
smaller binary split problems.

\subsection{Summary}
\label{sec:workload:summary}

Table~\ref{tab:workload_summary} summarizes the direct formulations and
their hardware bottlenecks. In every case, a direct QUBO formulation
exceeds the 45-spin budget at practical problem sizes. Pathfinding and
target sharing are primarily spin-limited, while tour construction is
limited by both spin count and coefficient range because one-hot
constraints require strong penalty terms. These observations motivate
the hardware-aware mapping pipeline and layer-specific Ising methods
described next.

\begin{table}[t]
\centering
\footnotesize
\caption{Direct QUBO formulations and hardware bottlenecks. Example
counts assume a $10{\times}10$ grid, 7--20 targets, and 3 robots with
12 targets.}
\label{tab:workload_summary}
\begin{tabular}{@{}lccc@{}}
\toprule
Layer & Scaling & Example & Hardware bottleneck \\
\midrule
Pathfinding & $O(|V|)$--$O(|V|T)$ & 100--2\,100 & spins + coeff.\ range \\
Tour construction & $O(N^2)$ & 49--400 & coeff.\ range + spins \\
Target sharing & $O(RN^2)$ & 432 & spin count \\
\bottomrule
\end{tabular}
\end{table}

%% file: sec/4_mapping_pipeline.tex
Building on the workload characterization in Section~\ref{sec:workload},
we propose a hardware-aware Ising mapping pipeline for compact
spin- and coefficient-limited CMOS Ising hardware. The pipeline
decomposes each planning layer into small Ising subproblems, generates
multiple hardware-compatible mappings, decodes and scores returned
candidates, and stitches accepted candidates into a planning decision.
The chip serves as a low-power candidate-generation accelerator, while
the host handles decomposition, validation, scoring, and global
consistency.

\subsection{Layer-Specific Decomposition}
\label{sec:mapping_pipeline:decomposition}

We first propose a layer-specific decomposition for each part of the
MRMT stack. The goal is not to solve the planning layer with a classical
heuristic and then polish the result with Ising hardware; instead, each
decomposition exposes small combinatorial subproblems for which the
Ising solver remains the candidate-generation engine.

For pathfinding, we propose a patch-sliding decomposition in which the
full map is replaced by a small spatial patch centered around the robot.
Only free cells inside this window are modeled, keeping the local
variable count below the spin limit.

For tour construction, we propose a clustered decomposition. The target
set is divided into distance-based clusters, the clusters are ordered at
a coarse level, and each cluster is solved as a local ordering
subproblem. The tour is assembled from Ising-generated local orders
rather than repaired by classical tour-improvement heuristics.

For target sharing, we propose a recursive split decomposition. Robots
and targets are recursively partitioned using distance-weighted graph
splits, and the sequence of Ising-generated splits produces robot-level
assignments.

Across layers, the design principle is the same: replace one large
hardware-incompatible optimization problem with a sequence of spin- and
coefficient-limited Ising subproblems. The host performs decomposition
and bookkeeping, but candidate solutions are generated by the Ising
solver.

\subsection{Multi-Mapping Generation}
\label{sec:mapping_pipeline:multimapping}

A central contribution of the pipeline is to treat hardware mapping as a
portfolio problem rather than as a single deterministic QUBO-to-Ising
conversion. A single conversion gives only one hardware realization of a
logical problem, and on a small physical chip that realization may be
poor. Rather than relying on one mapping rule, the pipeline generates
multiple independent, hardware-compatible variants of the same logical
problem. These variants are solved as a portfolio and can be dispatched
concurrently across solver cores or chips. Returned candidates are
decoded and selected using the original objective rather than the
distorted hardware coefficients.

The pipeline uses three forms of multi-mapping across the planning
layers.

\begin{figure}[t]
    \centering
    \includegraphics[width=\columnwidth]{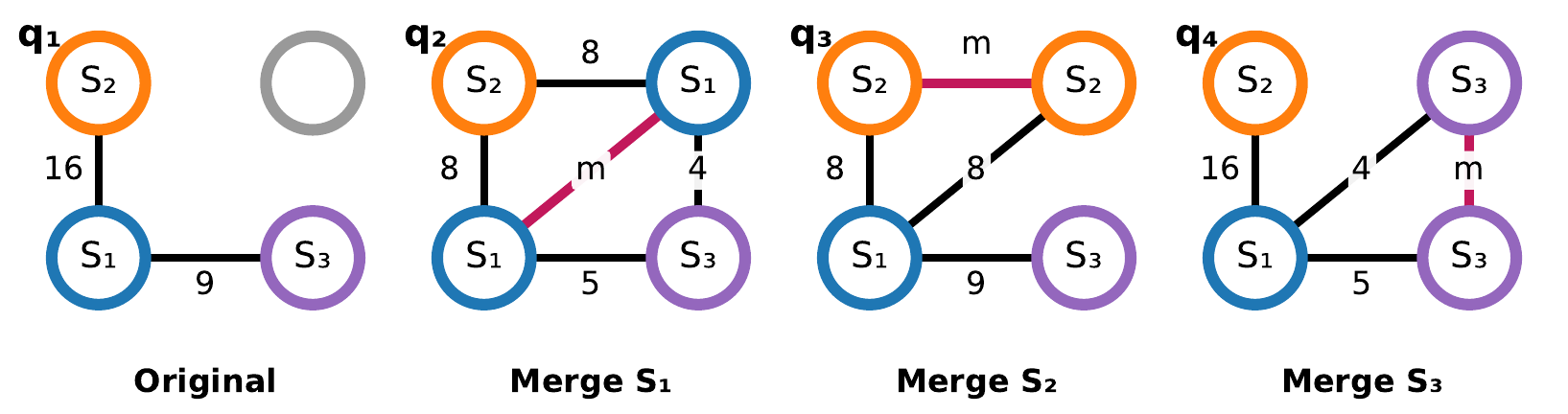}
    \vspace{-0.4cm}
    \caption{Representative spin-merging styles for mapping the same
    logical interaction pattern to hardware using an extra physical spin.
    Different merge choices redistribute the couplings differently,
    leading to different decoded outcomes.}
    \label{fig:merge_styles}
    \vspace{-0.4cm}
\end{figure}

\noindent\textbf{Spin merging.}
When individual Ising coefficients exceed the hardware range,
\emph{spin merging} duplicates an overloaded logical spin onto an
available extra physical spin and redistributes large couplings across
the original and duplicated copies. This reduces coefficient magnitudes
while approximately preserving the logical interaction structure. We
generate several spin-merging styles that choose the duplicated spin and
redistributed couplings differently, giving different
hardware-compatible views of the same patch problem. Figure~\ref{fig:merge_styles}
illustrates representative merge styles. Spin merging is used in
pathfinding, where local cell-based QUBOs can produce strong endpoint
and connectivity penalties.

\begin{figure}[t]
    \centering
    \includegraphics[width=\columnwidth]{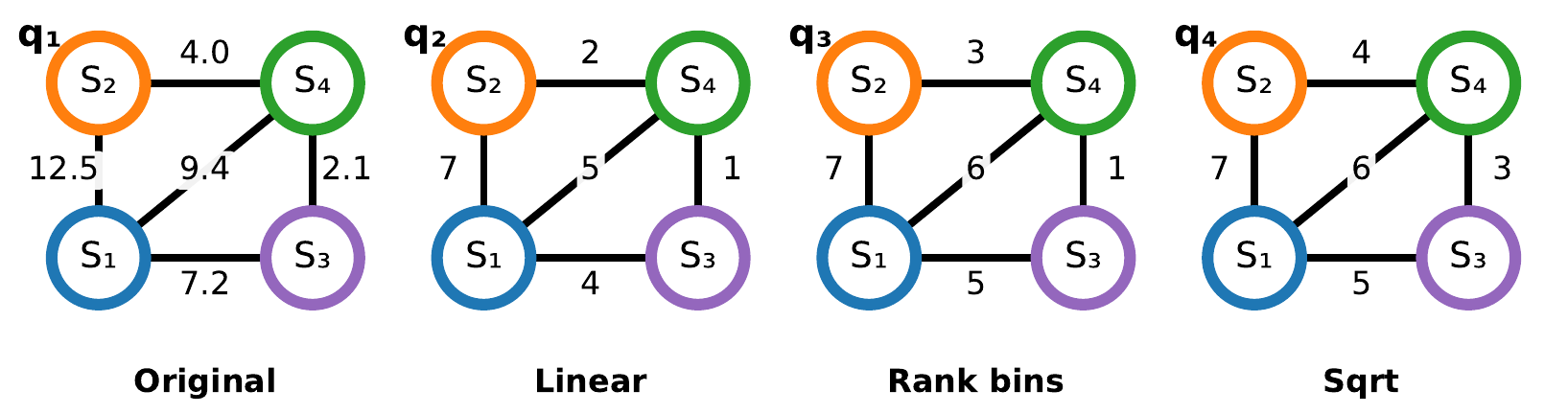}
    \vspace{-0.4cm}
    \caption{Representative quantized mappings of the same reduced Ising
    problem. The original floating-point couplers are shown on the left,
    followed by chip-compatible integer mappings produced by different
    quantization rules. Each mapping preserves a different view of the
    coefficient structure.}
    \label{fig:target_quantization}
    \vspace{-0.4cm}
\end{figure}

\noindent\textbf{Coefficient quantization.}
Distance-based objectives often produce real-valued Ising coefficients,
while the chip accepts integer coefficients. We therefore generate
multiple quantized mappings, including linear, clipped, rank-based, and
compressed mappings. These schemes preserve different aspects of the
coefficient structure, so they are evaluated as a portfolio rather than
selected by a fixed rule. Figure~\ref{fig:target_quantization} shows
representative quantized mappings. Quantization is used in target
sharing, where distance-weighted split objectives produce continuously
valued couplings.

\noindent\textbf{Spin-budget branching.}
Some logical subproblems are only slightly larger than the spin budget.
The pipeline then freezes a small number of spins and enumerates all
assignments of the frozen variables. For example, a 49-variable model
with a 45-spin budget requires freezing four spins, producing
$2^4 = 16$ reduced branches. The branches are solved independently, and
the candidate with the best original objective value is selected.
Branching is used in tour construction, where local TSP models can
reach 49 logical variables. Because this layer is limited by coefficient
range rather than only spin count, the reduced branches are evaluated
with the logical Ising solver in this work.

\subsection{Parallel Dispatch and Decode}
\label{sec:mapping_pipeline:dispatch}

The target hardware provides parallel solver resources, either as
multiple cores within a chip or as multiple chips on the same board. The
independent mappings generated by the portfolio can therefore be
evaluated concurrently. Recursive decomposition can also produce
independent subproblems at the same stage. In target sharing, for
example, multiple groups can be packed into one chip submission by
placing each split in a separate block of the Ising matrix and setting
all cross-block couplings to zero. The returned spin vector is then
decoded back into separate candidate splits.

After solver evaluation, each spin assignment is decoded into a
candidate planning decision: a local path segment, a target ordering, or
a group partition. Invalid candidates, such as disconnected paths,
violated one-hot constraints, or degenerate partitions, are rejected.
The remaining candidates are scored using the original unquantized
objective, and the best valid candidate is selected. This
decode-filter-score loop is always performed classically. Parallel
solver resources can reduce latency, but they do not change the
decoding and selection logic.

The host performs all surrounding planning logic and fallback decisions;
the Ising solver is used only for candidate generation within each
subproblem.

%% file: sec/5_case_studies.tex
This section presents three hardware-aware Ising-based planning methods
that instantiate the mapping pipeline from Section~\ref{sec:mapping_pipeline}.
Each method targets a different layer of the MRMT stack and is designed
around the spin-count, coefficient-range, and decode-validity constraints
of compact CMOS Ising hardware. Across all three methods, classical
logic decomposes, validates, scores, and stitches candidates, but does
not repair Ising outputs using traditional planning heuristics.
Table~\ref{tab:case_summary} summarizes the three methods.

\begin{table}[t]
\centering
\footnotesize
\caption{Summary of proposed Ising-based planning methods across the
three planning layers.}
\vspace{-0.2cm}
\label{tab:case_summary}
\begin{tabular}{@{}llll@{}}
\toprule
Layer & Decomposition & Mapping technique & Backend \\
\midrule
Pathfinding & spatial patches & spin merging & on-chip \\
Tour construction & clusters + patches & branching & logical Ising \\
Target sharing & recursive splits & quantization + batching & on-chip \\
\bottomrule
\end{tabular}
\vspace{-0.4cm}
\end{table}

\subsection{Pathfinding}
\label{sec:cases:pathfinding}

\begin{figure}[t]
    \centering
    \includegraphics[width=0.7\columnwidth]{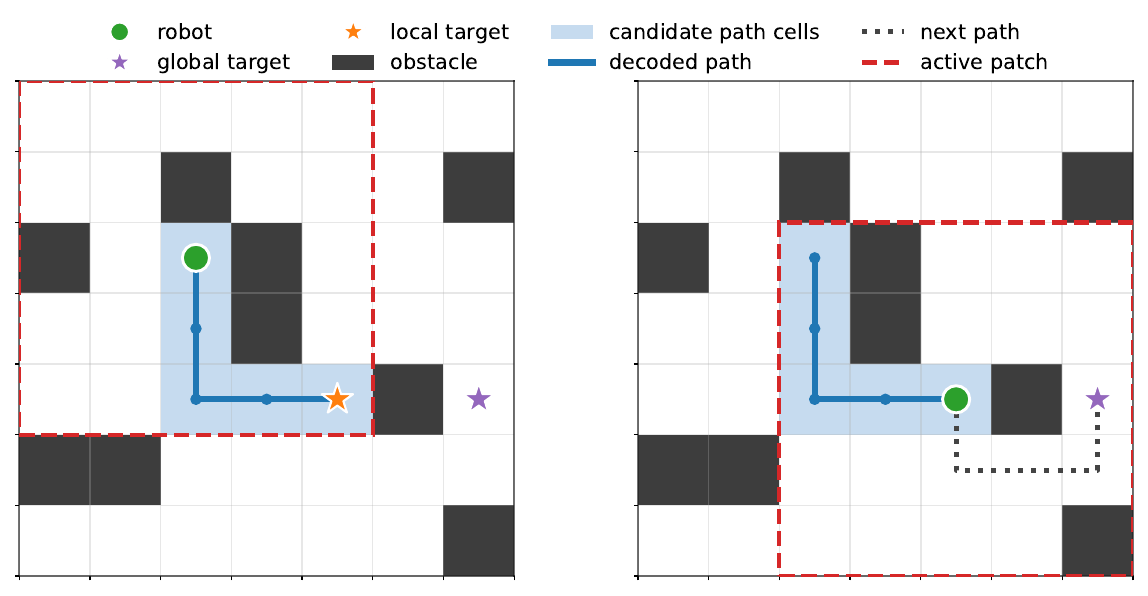}
    \vspace{-0.4cm}
    \caption{Overview of the patch-sliding pathfinding method. The left
    panel shows the current patch-level solve toward a local target. The
    right panel shows the committed segment and the continuation toward
    the final global target after the patch shifts forward.}
    \label{fig:system_overview}
    \vspace{-0.4cm}
\end{figure}

Rather than solving the full map as one Ising instance, we propose a
patch-sliding Ising pathfinder that uses the chip as a local accelerator
inside a receding-horizon planner. At each step, we center a small
$5{\times}5$ patch around the robot's current position. We model only
the free cells inside this local window and solve a local pathfinding
problem toward a local target on the patch boundary. Once a valid local
path is found, we append it to the global route, advance the robot, and
shift the patch forward. Figure~\ref{fig:system_overview} illustrates
this process.

\noindent\textbf{Local patch model.}
Within each patch, we use the cell-based formulation introduced in
Section~\ref{sec:workload:pathfinding}. We assign one binary variable to
each free cell and exclude obstacle cells from the local model. We pin
the robot position and the selected local target as fixed endpoints. The
local objective is
\begin{equation}
E_{\mathrm{patch}} =
A\,E_{\mathrm{end}}
+
A\,E_{\mathrm{conn}}
+
B \sum_{i \in \mathcal{P}} x_i,
\label{eq:patch}
\end{equation}
where $\mathcal{P}$ is the set of free cells in the current patch. In an
obstacle-free $5{\times}5$ patch, at most 23 active variables remain
after pinning the two endpoints. Obstacles reduce this number further,
so the patch-level problem fits within the 45-spin budget.

\noindent\textbf{Local-target ranking.}
The local target determines how the patch-level decision contributes to
the global route. We rank candidate boundary cells by Manhattan distance
to the final goal and try them in order, starting from the most
promising candidate. This ranking is performed classically as an
external heuristic rather than embedded into the Ising objective.
Keeping the ranking outside the chip avoids adding extra terms to the
QUBO and preserves the limited coefficient range for path-validity
penalties.

\begin{figure}[t]
    \centering
    \includegraphics[width=\columnwidth]{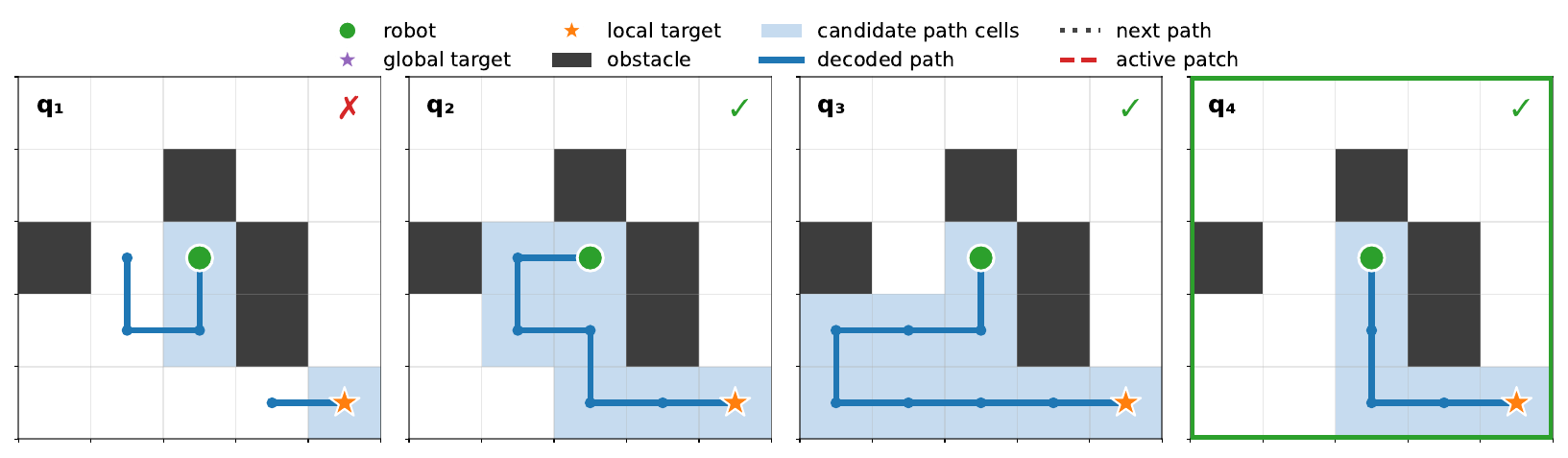}
    \vspace{-0.4cm}
    \caption{Representative decoded results from different hardware
    mappings of the same local patch problem. Invalid outputs are
    discarded, and the best valid local path is selected.}
    \label{fig:parallel_outputs}
    \vspace{-0.4cm}
\end{figure}

\noindent\textbf{Multi-mapping via spin merging.}
Although the patch model is small in variable count, its endpoint and
connectivity penalties can produce Ising coefficients outside the
hardware range after QUBO-to-Ising conversion. We address this using
spin merging, as described in Section~\ref{sec:mapping_pipeline:multimapping}.
For each patch problem, we generate several merge-style mappings and
solve them as a small portfolio. We decode the returned spin assignments
into candidate local paths, discard invalid candidates such as
disconnected selections or loops, and select the best remaining path
using the original local objective. Figure~\ref{fig:parallel_outputs}
shows a representative example in which different mappings of the same
patch produce different decoded outputs.

\noindent\textbf{Planning loop.}
Algorithm~\ref{alg:patch_sliding} summarizes the complete procedure.
The chip solves only the local patch problem; patch placement, candidate
ranking, decoding, validation, and path commitment are handled
classically.

\begin{algorithm}[t]
\small
\caption{Patch-sliding Ising pathfinding}
\label{alg:patch_sliding}
\begin{algorithmic}[1]
\State Initialize current position $p \gets s$
\While{$p \neq t$}
    \State Center a $5\times5$ patch around $p$
    \State Rank candidate local targets by Manhattan distance to $t$
    \For{candidate local target $g$ in ranked order}
        \State Build a local cell-based QUBO on the free patch cells
        \State Generate multiple merge-style hardware mappings
        \State Solve all mappings on the Ising chip
        \State Decode returned spins into candidate local paths
        \State Reject invalid local paths
        \If{at least one valid path exists}
            \State Select the best valid local path
            \State Append the selected segment to the global route
            \State Move $p$ to the end of the selected segment
            \State \textbf{break}
        \EndIf
    \EndFor
\EndWhile
\end{algorithmic}
\end{algorithm}

This pathfinding method runs on the physical chip. It exercises the
spin-merging and multi-mapping components of the pipeline and shows how
pathfinding can be made chip-compatible by reducing the full-map problem
to a sequence of small spatial patches.


\subsection{Tour Construction}
\label{sec:cases:tour}

We propose a clustered Ising tour-construction method for ordering the
targets assigned to one robot. A direct one-hot TSP formulation grows
quadratically with the number of targets and requires strong feasibility
penalties, as discussed in Section~\ref{sec:workload:tour}. Instead of
solving the full target-ordering problem at once, we decompose it into
smaller Ising subproblems through clustering, cluster ordering, local
TSP solving, and patch refinement. We then assemble the full tour from
these Ising-generated ordering decisions.

\noindent\textbf{Distance-Ising clustering.}
We first reduce the tour problem by recursively splitting the target set
with an Ising-compatible distance-weighted cut objective. Let $w_{ij}$
denote the obstacle-aware path distance between targets $i$ and $j$, and
let $x_i \in \{0,1\}$ indicate which side of the split target $i$
belongs to. For each split, we solve
\begin{equation}
\mathrm{cut}(x) =
\sum_{i<j}
w_{ij}
\left(
x_i + x_j - 2x_i x_j
\right).
\label{eq:tour_cut}
\end{equation}
Maximizing this objective separates distant targets, producing smaller
spatially coherent clusters that can be handled by spin-limited local
ordering subproblems. In our implementation, each local TSP subproblem
contains at most seven targets. We further refine the cluster set up to
an ordering limit of eight clusters. Since we fix the first cluster as
the one closest to the robot start, ordering $K$ clusters requires only
$(K{-}1)^2$ variables; with $K=8$, this gives 49 logical variables.

\begin{figure}[t]
    \centering
    \includegraphics[width=\columnwidth]{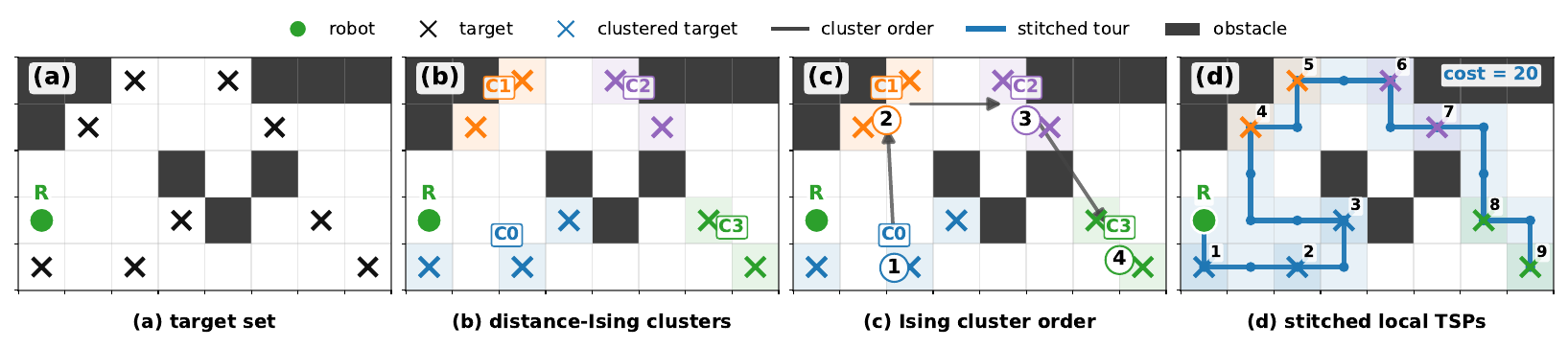}
    \vspace{-0.6cm}
    \caption{Overview of clustered Ising tour construction. Targets are
    recursively split into distance-based clusters, the clusters are
    ordered with an Ising/QUBO model, and each cluster is solved as a
    small local TSP before the local orders are stitched into a complete
    tour.}
    \label{fig:ising_cluster_pipeline}
    \vspace{-0.6cm}
\end{figure}

\noindent\textbf{Cluster ordering and local TSP solving.}
After clustering, we fix the cluster closest to the robot start as the
first cluster and order the remaining clusters with a position-indexed
QUBO using the same one-hot idea as the TSP formulation in
Section~\ref{sec:workload:tour}, but applied to clusters rather than
individual targets. We then solve each cluster as a local open TSP with
at most $7^2=49$ logical variables. The local model includes soft
boundary costs from the previous endpoint to the first local target and
from the last local target toward the next cluster. This lets the solver
choose the internal order and the entry/exit targets jointly, while
still encouraging the stitched route to connect well across cluster
boundaries. Figure~\ref{fig:ising_cluster_pipeline} illustrates the
clustering, ordering, and stitching process.

\noindent\textbf{Ising patch refinement.}
The stitched tour can still contain local mistakes, especially near
cluster boundaries. To improve the route while keeping the problem
small, we randomly select a contiguous patch of at most seven targets
and re-solve only that patch as a local one-hot TSP QUBO. The rest of
the tour is kept fixed. The patch objective includes the cost from the
previous waypoint into the patch, the internal ordering cost, and the
cost from the patch to the next waypoint. If the reordered patch
improves the full tour cost, we accept it; otherwise, we reject it. The
process stops after a fixed number of consecutive non-improving
attempts. No classical tour-improvement heuristic such as 2-opt or 3-opt
is used inside the proposed tour method; these algorithms appear only as
baselines in the evaluation.

\noindent\textbf{Spin-budget branching.}
The local ordering models can use up to 49 logical variables. With a
45-spin budget, we freeze four spins and enumerate all $2^4=16$
assignments of the frozen variables, as described in
Section~\ref{sec:mapping_pipeline:multimapping}. Each assignment
produces a reduced Ising problem with 45 spins. We solve the reduced
problems independently and select the decoded candidate with the best
original tour cost.

\noindent\textbf{Execution: logical Ising.}
Unlike pathfinding and target sharing, we evaluate the tour-construction
layer with a logical Ising solver rather than direct physical-chip
execution. The limitation is not only spin count but also coefficient
range: the one-hot permutation constraints require penalty strengths
that exceed the useful integer coefficient range of the chip. In
practice, we observe that direct chip samples often violate these
constraints and decode to invalid permutations. We therefore use a
logical Ising backend for this layer: Tabu sampling replaces the
physical chip as the Ising solver, while the decomposition, QUBO
encoding, branching, decoding, and scoring logic remain unchanged. This
method exposes an important hardware limitation: problem structure, not
just variable count, determines whether a subproblem can be executed on
the chip.

\noindent\textbf{Planning loop.}
Algorithm~\ref{alg:ising_cluster_tour} summarizes the complete procedure.

\begin{algorithm}[t]
\small
\caption{Clustered Ising tour construction}
\label{alg:ising_cluster_tour}
\begin{algorithmic}[1]
\State Compute start-to-target and target-to-target path distances
\State Initialize one cluster containing all targets
\While{some cluster is larger than the local TSP limit}
    \State Split the cluster using the distance-Ising objective
\EndWhile
\While{the number of clusters is below the ordering limit}
    \State Select a cluster for refinement
    \State Split it using the distance-Ising objective
\EndWhile
\State Fix the first cluster as the cluster closest to the robot start
\State Build an Ising/QUBO model to order the remaining clusters
\State Solve the cluster-ordering model
\State Initialize an empty global tour
\For{each cluster in the selected cluster order}
    \State Build a local one-hot TSP QUBO with soft boundary costs
    \State Solve the local TSP model
    \State Append the decoded local target order to the global tour
\EndFor
\State Initialize failed patch attempts $m \gets 0$
\While{$m < M$}
    \State Randomly select a contiguous local patch of the tour
    \State Solve the patch as a local one-hot TSP QUBO
    \If{the patched tour improves the full route cost}
        \State Accept the patch and set $m \gets 0$
    \Else
        \State Reject the patch and set $m \gets m+1$
    \EndIf
\EndWhile
\State \Return final target order
\end{algorithmic}
\end{algorithm}


\subsection{Target Sharing}
\label{sec:cases:target_sharing}

We propose recursive Ising target sharing for assigning targets to
robots. Instead of solving the full mTSP-style allocation and ordering
problem directly, we divide the global set of robots and targets into
smaller groups using recursive distance-weighted graph partitioning.
Each chip submission solves one or more small split problems, and the
sequence of Ising-generated splits gradually produces robot-target
assignments.

\begin{figure}[t]
    \centering
    \includegraphics[width=\columnwidth]{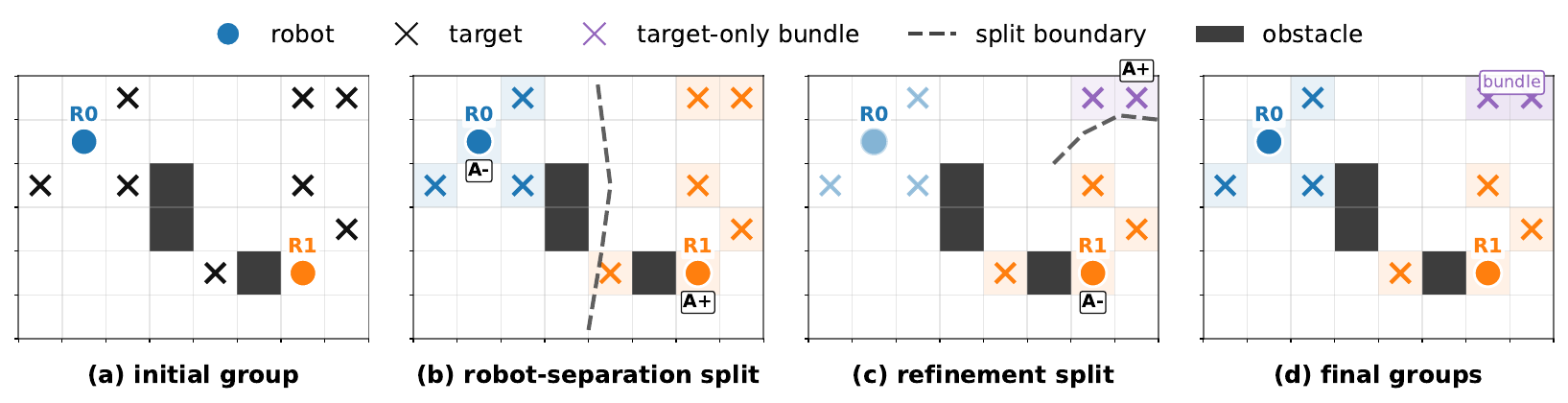}
    \vspace{-0.4cm}
    \caption{Recursive target sharing with anchored splits. The method
    starts from one group containing all robots and targets, then
    repeatedly divides groups using a distance-weighted split objective.
    Two anchors are fixed to opposite sides in each split, and refinement
    can create target-only bundles that are assigned later.}
    \label{fig:recursive_target_split}
    \vspace{-0.4cm}
\end{figure}

\noindent\textbf{Recursive split model.}
We represent robots and targets as nodes in a weighted graph whose edge
weights are pairwise path distances. Each Ising solve splits the current
group into two parts using the distance-weighted cut objective
\begin{equation}
\mathrm{cut}(x) =
\sum_{i<j}
w_{ij}
\left(
x_i + x_j - 2x_i x_j
\right),
\label{eq:target_cut}
\end{equation}
which encourages distant nodes to be placed on opposite sides. Starting
from one group containing all robots and targets, we repeatedly apply
this two-way split. We split groups containing more than one robot first,
so that robots are gradually separated. Once every group contains at most
one robot, the groups define a preliminary allocation. Additional
refinement splits can break large target groups into smaller bundles.
Figure~\ref{fig:recursive_target_split} illustrates this process.

\noindent\textbf{Anchor fixing.}
To guide each split, we fix two anchor nodes to opposite sides before
solving. If the group contains multiple robots, we choose the farthest
pair of robots. If the group contains one robot, we choose that robot
and the farthest target from it. If the group contains no robots, we
choose the farthest pair of targets. We substitute the anchors directly
into the QUBO rather than enforcing them with penalty terms. This
reduces the number of free variables and guarantees that the anchors
land on different sides of the split.

\noindent\textbf{Multi-mapping via quantization.}
After anchor substitution and QUBO-to-Ising conversion, we quantize the
coefficients to the chip's integer range using multiple quantization
rules, as described in Section~\ref{sec:mapping_pipeline:multimapping}.
We solve each quantized mapping on the chip and decode it into a
candidate partition. We score the candidates using the original
unquantized cut objective and select the best one.

\noindent\textbf{Batching independent splits.}
Recursive splitting can produce several independent split problems at
the same stage. We pack these splits into a single chip submission by
placing each split in a separate block of the Ising matrix with zero
cross-block couplings, as described in
Section~\ref{sec:mapping_pipeline:dispatch}. We then decode the returned
spin vector back into separate candidate splits and score each split
independently.

\noindent\textbf{Robotless bundle assignment.} 
After all splits are complete, groups containing exactly one robot
define that robot's base target set, while groups with no robot form
robotless bundles. We assign each bundle to the robot whose tour cost
increases the least when the bundle is added, using the downstream tour
solver for evaluation. The chip handles the partitioning stage; the
final bundle decision is made classically.

\noindent\textbf{Planning loop.}
Algorithm~\ref{alg:target_sharing} summarizes the complete procedure.

\begin{algorithm}[t]
\small
\caption{Recursive Ising target sharing}
\label{alg:target_sharing}
\begin{algorithmic}[1]
\State Build a node set from all robots and targets
\State Initialize one group containing all nodes
\While{some group contains more than one robot}
    \State Select groups that require splitting
    \For{each selected group}
        \State Choose two anchors inside the group
        \State Build the distance-weighted split QUBO
        \State Fix anchors to opposite sides by substitution
        \State Convert the reduced QUBO to native Ising form
        \State Generate multiple quantized chip mappings
    \EndFor
    \State Batch independent split problems when possible
    \State Solve the chip mappings and decode candidate splits
    \State Score candidates using the original split objective
    \State Replace each selected group by its best split
\EndWhile
\State Optionally refine large target groups into smaller bundles
\State Assign one-robot groups as base target sets
\State Assign robotless bundles by minimum marginal tour-cost increase
\end{algorithmic}
\end{algorithm}

The recursive split and quantization stages run on the physical chip.
This target-sharing method exercises the coefficient-quantization,
batching, and parallel-dispatch components of the pipeline and shows
that recursive graph partitioning is a natural fit for small Ising
hardware.

%% file: sec/6_evaluation.tex
This section evaluates the proposed hardware-aware Ising planning
pipeline. The evaluation asks three questions: whether each planning
layer produces useful solutions compared with classical baselines,
whether the complete Ising pipeline remains competitive end to end, and
what the mapping outcomes reveal about using a small spin- and
coefficient-limited Ising machine as a candidate-generation accelerator.

\subsection{Experimental Setup and Baselines}
\label{sec:evaluation:setup}

All experiments use randomly generated $10\times10$ grid worlds with
20\% known obstacle density and unit-cost four-neighbor motion. Robot starts, target locations, and obstacles are
sampled randomly. Obstacle-aware shortest-path distances are computed on
the grid and used by the tour-construction and target-sharing layers.
The number of robots, targets, and random instances varies by evaluation
layer and is stated in the corresponding result subsection.

\begin{figure}[t]
    \centering
    \includegraphics[width=0.7\linewidth]{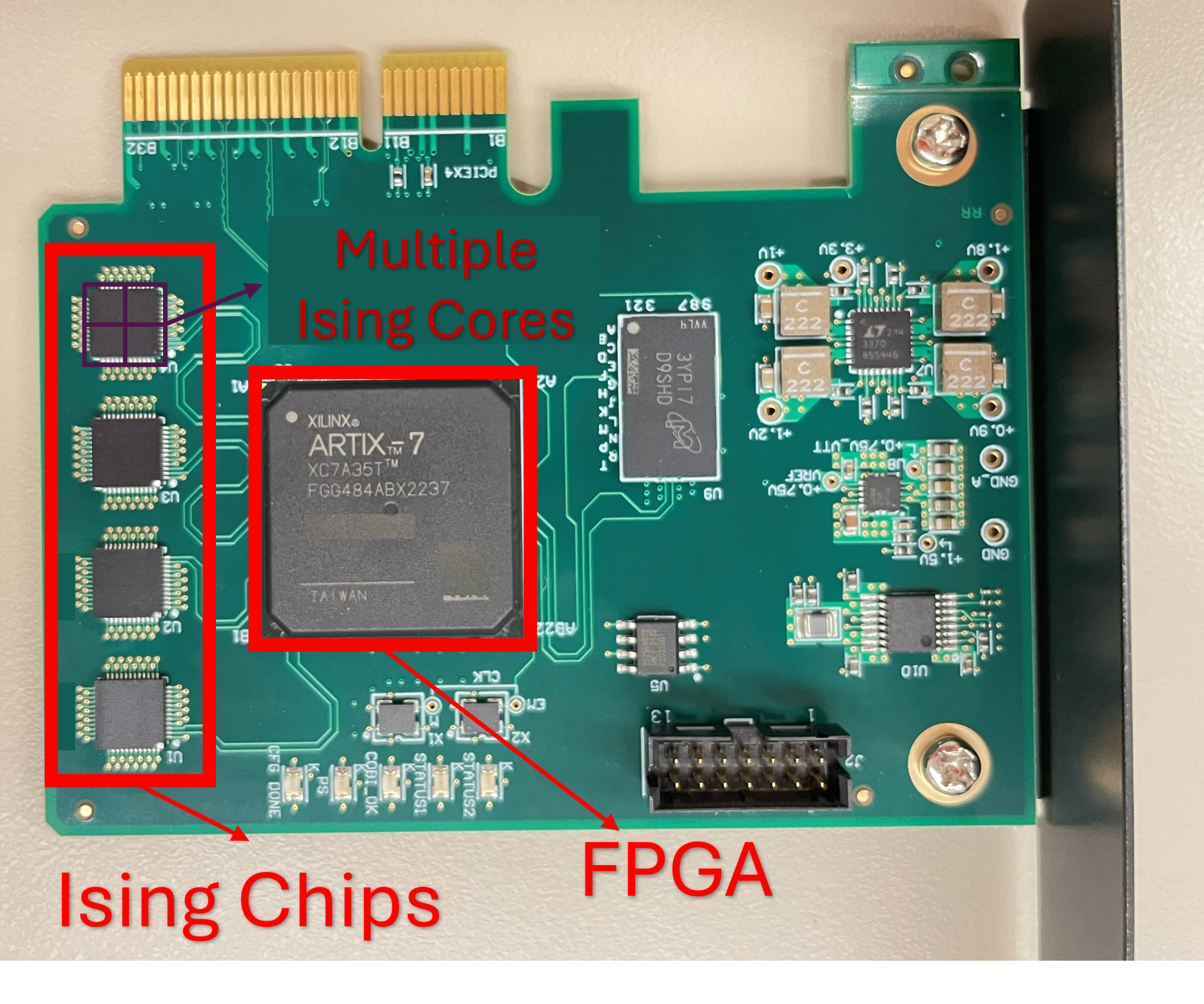}
    \vspace{-0.4cm}
    \caption{Experimental hardware setup. The CMOS Ising chip board is
    connected to a Linux host that performs decomposition, mapping,
    decoding, validation, and stitching.}
    \label{fig:chip_board}
    \vspace{-0.4cm}
\end{figure}

Physical-chip experiments use the CMOS Ising chip board shown
in Figure~\ref{fig:chip_board}. Each Ising core supports 45 all-to-all
connected spins with integer coefficients $h_i,J_{ij}\in[-7,+7]$. The
evaluation board contains four chips with two solver cores per chip, for
a total of eight parallel Ising cores. In this work, chip calls are
submitted sequentially to isolate per-call behavior and make energy
accounting unambiguous. We use 50~$\mu$s solve time and 9~mW chip power,
giving 0.45~$\mu$J per chip call. The chip is controlled by a Linux host
with an Intel Core i7-12700 CPU and 16~GiB of memory, running Ubuntu
24.04 and Python 3.12. Classical baselines and all host-side planning
logic run on the same host.

For classical baselines, solver time is measured as CPU wall-clock
runtime, and the CPU energy is computed using a 15~W per-core
power estimate obtained by normalizing the CPU TDP by the number of
physical cores. For physical-chip layers, chip time and chip energy are
computed from the number of chip calls using 50~$\mu$s and
0.45~$\mu$J per call. For tour construction, the current chip coefficient
range is insufficient for the one-hot permutation constraints, so the
clustered Ising tour method is evaluated with spin-limited logical Tabu
sampling; its reported chip time and energy are projections obtained by
replacing each logical Tabu call with one chip solve. Reported Ising energy is accelerator energy: measured chip-call
energy for physical-chip layers and projected chip-call energy for
logical-Ising tour construction.

For pathfinding, we compare against breadth-first search (BFS), A*, and
Greedy Best-First Search (GBFS). For tour construction, we compare
against nearest-neighbor ordering, 2-opt, 3-opt, and exact brute force
when the target count permits. For target sharing, we compare against
round-robin assignment, parallel single-item auction (PSA), and
sequential single-item auction (SSA). The end-to-end classical pipelines
combine A* pathfinding with either PSA and nearest-neighbor ordering
(fast classical) or SSA and 3-opt (strong classical). The end-to-end
Ising pipeline combines recursive Ising target sharing, clustered
logical-Ising tour construction, and patch-sliding Ising pathfinding.
These classical methods are used only as baselines; the proposed Ising
methods do not invoke A*, 2-opt, 3-opt, auctions, or other classical
repair heuristics to repair Ising outputs.

We report success rate, solution cost, solver time, and energy proxy.
Cost is path length for pathfinding, optimality gap or tour length for
tour construction, and total route length for target sharing and
end-to-end planning.

\subsection{Results}
\label{sec:evaluation:results}

We present results separately for the three planning layers and then
evaluate the complete pipeline. Each plot reports distributions over the
random instances described in the corresponding subsection. Boxplots show
the median, interquartile range, and outliers. For cost, time, and
energy, lower is better; for success rate and valid decode rate, higher
is better.

\subsubsection{\textbf{Pathfinding}}
\label{sec:evaluation:pathfinding}

\begin{figure}[t]
    \centering
    \includegraphics[width=\columnwidth]{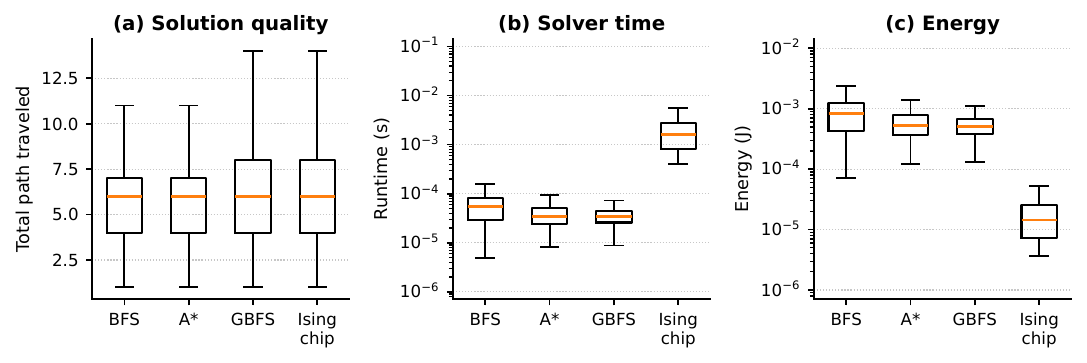}
    \vspace{-0.7cm}
    \caption{Pathfinding evaluation on 1000 random single-target
    queries. Quality is measured by total path length. Classical methods
    report CPU runtime and core-normalized CPU energy proxy; the Ising
    method reports accumulated chip time and chip energy.}
    \label{fig:eval_pathfinding}
    \vspace{-0.7cm}
\end{figure}

Figure~\ref{fig:eval_pathfinding} compares the patch-sliding Ising
pathfinder against BFS, A*, and GBFS on 1000 random single-target
queries. All classical methods solve every instance. The Ising
pathfinder solves 996 out of 1000 instances despite using probabilistic
local patch solves. Its median path
length is 6.0, matching the medians of BFS, A*, and GBFS.

The Ising pathfinder uses a median of 14~$\mu$J per query, including
every hardware call, compared with 523~$\mu$J for A*, corresponding to a
roughly 37$\times$ energy reduction. This comes at the cost of latency:
the Ising pathfinder accumulates a median of 1.6~ms in chip annealing
time, corresponding to 32 chip calls at 50~$\mu$s per call, whereas A*
completes in 35~$\mu$s on the CPU.

\subsubsection{\textbf{Tour Construction}}
\label{sec:evaluation:tour}

\begin{figure}[t]
    \centering
    \includegraphics[width=\columnwidth]{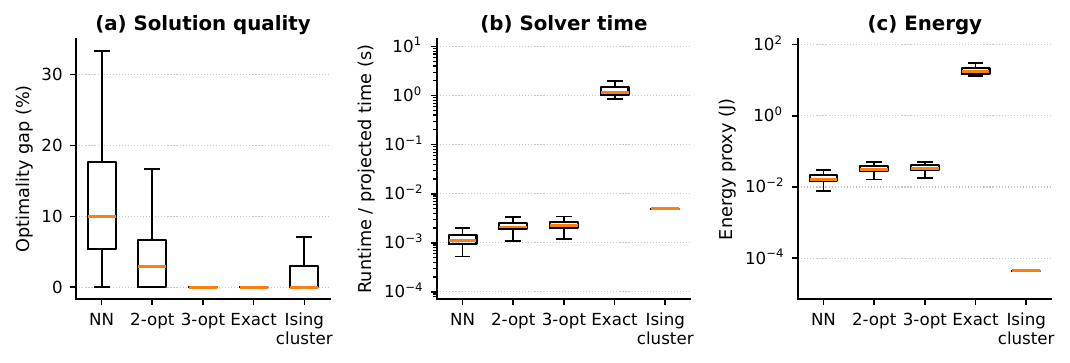}
    \vspace{-0.7cm}
    \caption{Tour-construction evaluation on 100 random 10-target
    instances. Quality is reported as optimality gap relative to exact
    brute force. The clustered Ising tour method uses a 45-spin-limited
    logical Tabu backend; its time and energy are projected by replacing
    each Tabu call with one chip solve.}
    \label{fig:eval_tour}
    \vspace{-0.4cm}
\end{figure}

Figure~\ref{fig:eval_tour} evaluates the tour-construction layer on 100
random 10-target instances. Exact brute force provides the optimal
reference, while nearest neighbor, 2-opt, and 3-opt provide classical
heuristic baselines.

Tour construction is evaluated with a logical Ising backend rather than
direct physical-chip execution. The clustered Ising formulation works
under logical Ising evaluation, but the one-hot permutation constraints
require penalty coefficients larger than the useful chip coefficient
range. To isolate the algorithmic value of the formulation, we use a
spin-limited Tabu backend with the same 45-spin subproblem limit as the
chip, but without the chip's coefficient-range restriction.

The clustered Ising tour method achieves a median optimality gap of
0.0\%, matching exact brute force and 3-opt. Its mean optimality gap is
1.63\%, close to the 1.31\% mean gap of 3-opt and better than nearest
neighbor and 2-opt, which have mean gaps of 11.61\% and 4.62\%,
respectively. For projected time and energy, the clustered Ising tour
method uses a median of 98 logical Ising calls, corresponding to
4.9~ms projected chip time and 44~$\mu$J projected chip energy. This
projected energy is roughly 770$\times$ lower than the core-normalized
CPU energy proxy of 3-opt.

\subsubsection{\textbf{Target Sharing}}
\label{sec:evaluation:target_sharing}

\begin{figure}[t]
    \centering
    \includegraphics[width=\columnwidth]{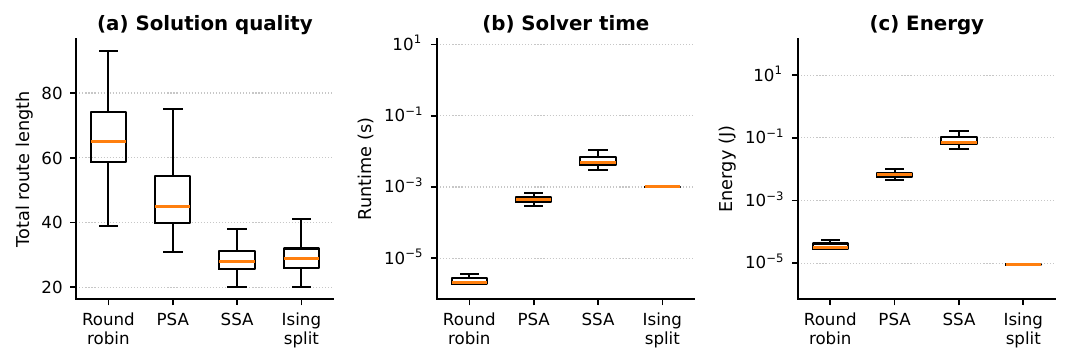}
    \vspace{-0.7cm}
    \caption{Target-sharing evaluation on 100 random instances with
    3 robots and 10 targets. Quality is reported as total route length.
    Classical methods report CPU runtime and energy proxy; recursive
    Ising target sharing reports accumulated chip time and chip energy.}
    \label{fig:eval_target_sharing}
    \vspace{-0.4cm}
\end{figure}

Figure~\ref{fig:eval_target_sharing} evaluates recursive Ising target
sharing on 100 random instances with 3 robots and 10 targets. Round
robin, PSA, and SSA provide classical allocation baselines. Our proposed recursive Ising target-sharing method maps naturally to the
physical chip because each split is a max-cut-like binary partitioning
problem, so every spin assignment corresponds to a valid split candidate.

In allocation quality, the recursive Ising target-sharing method is
close to the strongest classical baseline. SSA achieves the best median
total route cost of 28.0, while recursive Ising target sharing obtains
29.0. Relative to the best observed result per instance, recursive
Ising target sharing has a median gap of 2.78\%, compared with 0.0\%
for SSA. Both methods produce balanced allocations, with identical
median makespan of 16.0; recursive Ising target sharing also has a
slightly lower median workload imbalance, 12.0 compared with 14.0 for
SSA.

The energy comparison is strongest in this layer. Recursive Ising target
sharing uses a median of 20 chip calls organized into 5 batches,
corresponding to 1.0~ms accumulated chip annealing time and 9~$\mu$J
chip energy. SSA uses a median of 73~mJ under the core-normalized CPU
energy proxy, giving a roughly 8{,}000$\times$ energy reduction.

\subsubsection{\textbf{End-to-End Pipeline}}
\label{sec:evaluation:endtoend}

\begin{figure}[t]
    \centering
    \includegraphics[width=\columnwidth]{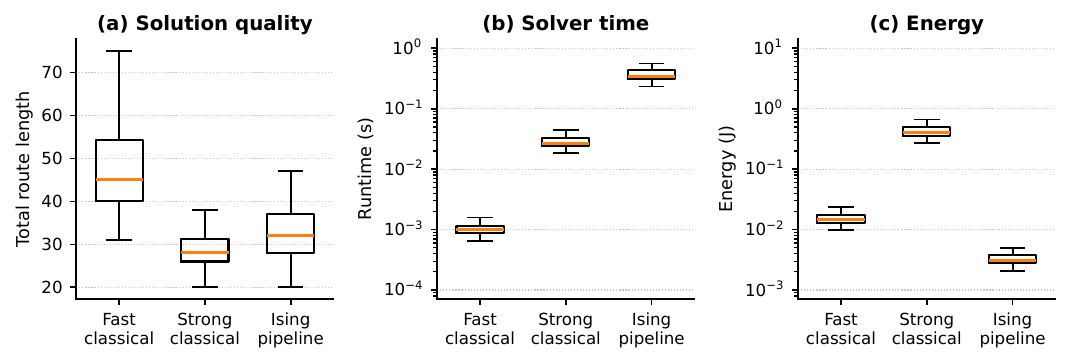}
    \vspace{-0.7cm}
    \caption{End-to-end pipeline comparison on instances with 3 robots
    and 10 targets. The Ising pipeline uses physical-chip calls for
    pathfinding and target sharing, and projected chip costs for
    clustered Ising tour construction using logical Tabu sampling.}
    \label{fig:eval_endtoend}
    \vspace{-0.6cm}
\end{figure}

Figure~\ref{fig:eval_endtoend} compares the full Ising pipeline against
two classical pipelines on random multi-robot multi-target instances
with 3 robots and 10 targets. The fast classical pipeline uses PSA,
nearest-neighbor tour construction, and A*. The strong classical
pipeline uses SSA, 3-opt, and A*. The Ising pipeline uses physical-chip
Ising calls for target sharing and pathfinding, and logical Tabu
sampling for tour construction.

The Ising pipeline achieves a median total route cost of 32.0, compared
with 28.0 for the strong classical pipeline and 45.0 for the fast
classical pipeline. Its median gap to the strong classical result is
9.1\%. Its median energy is 3.07~mJ, compared with 399~mJ for the strong
classical pipeline and 14.8~mJ for the fast classical pipeline,
corresponding to roughly 130$\times$ and 4.8$\times$ lower energy,
respectively.

The energy advantage comes at the cost of higher latency. The Ising
pipeline takes a median of 341~ms, compared with 27~ms for the strong
classical pipeline and 0.99~ms for the fast classical pipeline. The
bottleneck is pathfinding: completed Ising-pipeline runs use a median of
6{,}720 pathfinding chip calls, corresponding to 336~ms at 50~$\mu$s per
call. Recursive Ising target sharing uses only 20 chip calls, and
clustered Ising tour construction contributes 88 projected Ising calls.

\subsubsection{\textbf{Multi-Mapping Outcomes}}
\label{sec:evaluation:multimapping}

Table~\ref{tab:mapping_outcomes} tests our multi-mapping hypothesis:
hardware-compatible mapping should be treated as a portfolio rather than
as a fixed rule. For pathfinding, we generate four spin-merging styles
for each local patch problem. Each style alone decodes to a valid local
path only about 45\% of the time, but the failures occur on different
patch instances. Using the four styles together lets the pathfinder solve
996 out of 1000 queries. The selected-solution win shares are nearly
uniform, showing that no single spin-merging style dominates.

For target sharing, each recursive split is mapped using four
coefficient-quantization rules. All four quantization schemes produce
valid candidates, and their win shares are again approximately uniform,
ranging from 24\% to 25\%. These results support the multi-mapping
strategy: no single mapping rule dominates, so evaluating multiple
hardware-compatible variants improves robustness without committing to a
fixed mapping heuristic. Parallel solver resources could evaluate these
independent mappings concurrently, although this work submits them
sequentially for unambiguous chip-call accounting.

\begin{table}[t]
\centering
\footnotesize
\caption{Mapping outcome summary for the two physical-chip layers.
Pathfinding reports valid-decode rate and selected-solution win share
for each spin-merging style. Target sharing reports candidate share and
selected-solution win share for each coefficient-quantization rule.}
\vspace{-0.3cm}
\label{tab:mapping_outcomes}
\begin{tabular}{@{}llcc@{}}
\toprule
Layer & Mapping choice & Valid/candidate share & Win share \\
\midrule
Pathfinding & Overload & 44.8\% valid & 24.9\% \\
Pathfinding & Max-coeff. & 45.1\% valid & 25.1\% \\
Pathfinding & High-degree & 45.1\% valid & 25.0\% \\
Pathfinding & Balanced & 45.1\% valid & 25.0\% \\
\midrule
Target sharing & Linear & 25.0\% candidates & 25.2\% \\
Target sharing & Clip95 & 25.0\% candidates & 25.4\% \\
Target sharing & Rank & 25.0\% candidates & 25.3\% \\
Target sharing & Sqrt & 25.0\% candidates & 24.0\% \\
\bottomrule
\end{tabular}
\vspace{-0.3cm}
\end{table}

%% file: sec/7_related_work.tex
QUBO and Ising formulations provide a common interface for mapping
combinatorial optimization problems to specialized hardware. Lucas
surveys Ising formulations for many NP-hard problems, while Mohseni
et al. review Ising machines as hardware solvers for combinatorial
optimization~\cite{lucas2014ising,mohseni2022ising}. Hardware
implementations span CMOS annealing processors, fully connected digital
annealers, probabilistic p-bit machines, coupled-oscillator Ising chips,
and adaptive ReRAM-based solvers~\cite{yamaoka2016ising,
takemoto2020ising,yamamoto2021statica,aadit2022massively,
aadit2024alltoall,cilasun2025cobi,chiang2024reaim}. These works
demonstrate that Ising-style optimization can be realized in hardware.
Our work asks a different question: how a compact, low-power CMOS Ising
chip can be used inside a robotics planning stack under real spin-count,
coefficient-range, parallelism, and decode-validity constraints.

Several works study QUBO or Ising formulations for planning, routing,
and scheduling problems~\cite{lucas2014ising,jaroszewski2020routing,
zhang2022qubojsp}. These efforts show that planning and scheduling
problems can often be written in QUBO form, but direct formulations can
be too large, too penalty-sensitive, or too dependent on postprocessing
for resource-limited hardware. In contrast, we study a multi-layer
robotics workload and propose hardware-aware Ising methods for target
sharing, tour construction, and pathfinding. The proposed methods use
Ising solvers for candidate generation, while classical logic validates,
scores, and stitches candidates rather than repairing them with
traditional planning heuristics.

A broader architecture literature studies how applications should be
mapped to specialized accelerators~\cite{hegde2021mind,huang2021cosa}.
These works are not Ising-specific, but they support the same
systems-level lesson: accelerator efficiency depends on mapping choices,
not only on the hardware. We propose a multi-mapping pipeline for
approximate Ising solvers that uses decomposition, spin merging,
coefficient quantization, spin-budget branching, decode-time validation,
and fallback behavior. Our evaluation shows that no single merge or
quantization rule dominates, motivating the use of mapping portfolios.

Hardware acceleration for robotics and autonomous systems is another
closely related area~\cite{murray2016microarchitecture,
murray2019programmable,liu2021archytas,shah2023energy,hao2023blitzcrank,
hao2024orianna,yang2023dadu,niu2024fpga,huang2025corki}. These systems
primarily target deterministic numerical kernels, collision checking,
factor-graph inference, quadratic programming, dynamics, or control. Our
work targets a different role for hardware: using an Ising chip as a
low-power candidate generator for small combinatorial planning
subproblems in a larger classical planning loop.

The planning workload itself builds on classical multi-robot planning
and routing. Target sharing is related to multi-robot task allocation
and the multiple traveling salesman problem~\cite{gerkey2004formal,
korsah2013comprehensive,quinton2023market,chakraa2023mrta,
bektas2006multiple}. Tour construction is related to TSP-style target
ordering, while pathfinding builds on shortest-path methods such as
Dijkstra's algorithm and A*~\cite{dijkstra1959note,hart1968formal}.
Multi-agent path finding also studies path planning for multiple robots,
often through search or compilation to SAT, CSP, or MILP~\cite{
surynek2022mapf}. We use these classical methods as baselines and
global orchestration tools, while the proposed local combinatorial cores
use Ising-based candidate generation.

%% file: sec/8_conclusion.tex
This paper studied whether compact CMOS Ising machines can be used as
practical accelerators for multi-robot multi-target planning. Rather
than treating the Ising chip as a standalone optimizer, we use it as a
candidate-generation accelerator inside a classical planning loop. This
framing is important because realistic planning instances exceed current
hardware limits in several ways: pathfinding and target sharing quickly
exceed the spin budget, while tour construction stresses the available
coefficient range through one-hot permutation constraints.

We characterized three planning layers---target sharing, tour
construction, and pathfinding---and showed that each layer exposes a
different hardware bottleneck. Based on this characterization, we
proposed a hardware-aware Ising mapping pipeline that decomposes global
planning problems into chip-sized subproblems, generates multiple
hardware-compatible mappings, validates decoded outputs, and stitches
accepted candidates into a global plan. We instantiated the pipeline with
three Ising-based planning methods: patch-sliding Ising pathfinding,
clustered Ising tour construction, and recursive Ising target sharing.

The main lesson is that current compact Ising chips are useful only when
the surrounding system is designed around their constraints. Spin count,
coefficient range, quantization, and decode validity are not
implementation details; they determine which parts of the planning stack
can run on hardware and which must remain logical or classical. Our
results show that pathfinding and recursive target sharing can be mapped
to physical-chip subproblems, while one-hot tour construction requires
logical Ising sampling because the physical coefficient range is
insufficient for reliable permutation constraints.

More broadly, this work suggests that near-term Ising acceleration for
robotics should focus on hardware-aware candidate generation rather than
global replacement of classical planners. Classical logic should manage
decomposition, validation, scoring, fallback, and global consistency,
while Ising hardware generates candidates for carefully selected local
combinatorial subproblems. As CMOS Ising chips scale in spin count,
coefficient precision, and programmability, the proposed pipeline can be
extended to larger patches, richer assignment models, and more tightly
integrated multi-robot planning systems.